\documentclass[11pt]{article}
\usepackage[pdftex]{graphicx,color} 
\usepackage{jheppub}
\usepackage{amsmath}
\usepackage{amssymb}
\usepackage{comment}
\usepackage{multirow}
\usepackage{mathtools}
\usepackage{dsdshorthand}
\usepackage{tikz}
\usepackage{shuffle}
\usepackage{subcaption}
\usetikzlibrary{arrows.meta}
\newcommand{\bea}{\begin{equation}\begin{aligned}}
\newcommand{\eea}[1]{\label{#1}\end{aligned}\end{equation}}
\newcommand{\beq}{\begin{equation}}
\newcommand{\eeq}{\end{equation}}

\definecolor{integration_contours}{rgb}{0.7,0.0,0.13}
\definecolor{graph_poles}{rgb}{0.25, 0.35, 1.0}

\title{On the Regge behaviour of the AdS Virasoro-Shapiro Amplitude}
\author[a]{Luis F. Alday,}
\author[a]{Maria Nocchi,}
\author[a]{Cl\'ement Virally,}
\author[b]{Xinan Zhou}
\affiliation[a]{Mathematical Institute, University of Oxford, Andrew Wiles Building, Radcliffe Observatory Quarter, Woodstock Road, Oxford, OX2 6GG, U.K.}
\affiliation[b]{Kavli Institute for Theoretical Sciences, University of Chinese Academy of Sciences, Beijing 100190, China.}

\abstract{
We compute the AdS Virasoro-Shapiro amplitude in the Regge limit, in terms of the CFT data of the exchanged operators in the leading Regge trajectory. To any order in the small curvature expansion, the result can be written in terms of derivatives of the flat space Virasoro-Shapiro amplitude in the Regge limit. The result also admits an integral representation involving single-valued logarithms, fully consistent with recent proposals for the full AdS Virasoro-Shapiro amplitude. 
}

\emailAdd{alday@maths.ox.ac.uk, nocchi@maths.ox.ac.uk, clement.virally@maths.ox.ac.uk, xinan.zhou@ucas.ac.cn}

\begin{document}

\maketitle
\section{Introduction}

Over the last couple of years, substantial progress has been made in the computation of the four-point scattering amplitude of gravitons in type IIB string theory on $AdS_5 \times S^5$ at tree level \cite{Alday:2022uxp,Alday:2022xwz,Alday:2023jdk,Alday:2023mvu}. Due to the presence of Ramond-Ramond fluxes, the usual world-sheet techniques, such as the Ramond–Neveu–Schwarz formalism \cite{Polchinski:1998rr}, do not apply to this case, and other formalisms, such as the Green-Schwarz \cite{Green:1980zg} and pure spinor \cite{Berkovits:2000fe}, are not yet developed to the point of providing scattering amplitudes in this background.  On the other hand, for the special case of the four graviton amplitude on $AdS_5 \times S^5$ at tree level, also denoted as the AdS Virasoro-Shapiro amplitude $A(S,T)$, we have indirect tools that have led to the aforementioned progress. 

The AdS/CFT duality relates this amplitude to the planar four-point correlator of stress tensors in ${\cal N}=4$ Super-Yang-Mills (SYM). In particular, this relates the analytic structure of $A(S,T)$, namely its residues and poles, to the CFT data of intermediate operators \cite{Alday:2022uxp}. This CFT data can in turn be computed, at least in principle, by the powerful integrability techniques developed for planar ${\cal N}=4$ SYM, see {\it e.g.} \cite{Gromov:2011de,Basso:2011rs,Gromov:2011bz}. The analytic structure of $A(S,T)$ turns out to be very constraining when supplemented by a conjectural single-valuedness property, according to which the low energy expansion of the AdS Virasoro-Shapiro amplitude contains only single-valued zeta values \cite{Alday:2022xwz}. 

From these developments, a `world-sheet' picture emerged \cite{Alday:2023jdk,Alday:2023mvu}. In a small curvature expansion, the AdS Virasoro-Shapiro amplitude admits a representation in terms of a genus zero integral:
\begin{equation}
A(S,T) = \int d^2z |z|^{-2S-2}|1-z|^{-2T-2}G(S,T,z) \ ,
\end{equation}
where $G(S,T,z)= \sum \frac{1}{R^{2k}}G^{(k)}(S,T,z)$, with $R$ the common radius of $AdS_5$ and $S^5$ and $G^{(0)}(S,T,z)=1/(S+T)^2$ the usual insertion in the case of the flat space Virasoro-Shapiro amplitude. More generally, $G^{(k)}(S,T,z)$ is a rational function in $S,T$ and a transcendental single-valued function of weight $3k$ in $z$. Combining this structure with the dimension of the Konishi operator in planar ${\cal N}=4$ SYM, $G^{(1)}(S,T,z)$ and $G^{(2)}(S,T,z)$ were also fully determined\footnote{Modulo certain ambiguities that integrate to zero.}. These results pass several highly non-trivial checks, such as reproducing all localisation results, see {\it e.g.} \cite{Chester:2020dja}, and all CFT data available from integrability \cite{Gromov:2011de,Basso:2011rs,Gromov:2011bz}. 

 Towards the formidable task of determining $A(S,T)$ to all orders in $1/R$, we can focus on more manageable limits. One such limit is the high energy limit, where to each order in $1/R$ we only keep the leading large $S,T$ contribution. In this limit, $A(S,T)$ can be determined to all orders \cite{Alday:2023pzu}. In this paper, we consider a much richer limit, namely the large $T$, finite $S$ Regge regime. In the Regge limit, the amplitude encodes full information on the intermediate operators in the leading Regge trajectory. In the AdS/CFT context, these are stringy/short operators, for instance, the Konishi operator. Since their CFT data is a prime target of integrability, this is a very interesting limit to consider, with great potential synergy between integrability developments and the program of computing the AdS Virasoro-Shapiro amplitude.\footnote{See also \cite{Caron-Huot:2022sdy,Cavaglia:2022yvv,Julius:2023hre,Chester:2023ehi,Ekhammar:2024rfj} for very interesting developments on the interplay between integrability and the AdS Virasoro-Shapiro amplitude.}

In the first part of this paper, we determine the AdS Virasoro-Shapiro amplitude in the Regge limit, $A_{Regge}(S,T)$, in terms of the CFT data of the exchanged operators in the leading Regge trajectory. In some sense, this can be seen as the Borel transform of the results of \cite{Costa:2012cb}, but the translation is quite involved, due to the complicated analytic structure of $A(S,T)$. In the second part of the paper, we work out the implications of this result at the level of the world-sheet representation for the AdS Virasoro-Shapiro amplitude. In particular, we obtain a functional equation that determines the integrand -- in the appropriate Regge limit -- to all orders in $1/R$, up to ambiguities that integrate to zero in this limit. Our result is fully consistent with, and gives strong evidence for, the proposal put forward in \cite{Alday:2022xwz,Alday:2023jdk,Alday:2023mvu}. We defer some technical details to the appendices. Appendix~\ref{app:exchanges} includes details on the exchange amplitude, while Appendix~\ref{app:transforms} includes details on the several integral transforms used in the paper. In Appendix~\ref{app:propagator}, we show that the analytic structure of the AdS Virasoro-Shapiro amplitude in a $1/R$ expansion mimics that of the AdS bulk to bulk propagator in `momentum space', when expanded around flat space.

\section{Regge limit of the AdS VS amplitude}
We are interested in the Regge behaviour of the AdS Virasoro-Shapiro (VS) amplitude. The AdS VS amplitude is defined in terms of the four-point correlator of chiral primary operators in ${\cal N}=4$ SYM at leading non-trivial order in the large central charge expansion:
\begin{equation}
\left. \langle {\cal O}^{I_1J_1}(x_1) {\cal O}^{I_2J_2}(x_2) {\cal O}^{I_3J_3}(x_3) {\cal O}^{I_4J_4}(x_4) \rangle \right|_{\frac{1}{c}} = G^{I_i J_i}(x_i)_{tree} \ .
\end{equation}
Here ${\cal O}^{I_1J_1}(x)$ is the superconformal primary operator of the stress tensor multiplet. It is a scalar and has protected dimension $\Delta=2$.  The indices $I,J$ are in the fundamental representation of the $SO(6)$ R-symmetry group and the operator transforms in the symmetric traceless representation of rank 2. Superconformal Ward identities fix the dependence on the R-symmetry indices through a computable prefactor:
\begin{equation}
G^{I_i J_i}(x_i)_{tree} = factor(I_i,J_i,x_i) \times {\cal T}(U,V) \ .
\end{equation}
The non-trivial dynamics is encoded in the reduced correlator ${\cal T}(U,V)$, which is a function of the two conformally invariant cross-ratios 
\begin{equation}
U= \frac{x_{12}^2 x_{34}^2}{x_{13}^2 x_{24}^2} \ ,~~~~~V= \frac{x_{14}^2 x_{23}^2}{x_{13}^2 x_{24}^2} \ .
\end{equation}
The reduced correlator ${\cal T}(U,V)$ is related to the AdS Virasoro-Shapiro amplitude by a chain of integral transforms. First, we introduce the Mellin transform\footnote{Defined this way, $M(s, t)$ is usually denoted as the {\it reduced} Mellin amplitude, since it is the transform of the reduced correlator. Since this is the only Mellin amplitude we will deal with in this paper, we simply call it the Mellin amplitude.}, in the conventions of~\cite{Alday:2023mvu}:
\beq
  \cT(U, V)
   = \int_{-i \infty}^{i \infty} \frac{ds  dt}{(4 \pi i)^2} U^{\frac{s}{2}+\frac23} V^{\frac{t}{2} - \frac43}
    \Gamma \bigg(\frac43 - \frac{s}{2} \bigg)^2 \Gamma \bigg(\frac43 - \frac{t}{2} \bigg)^2 \Gamma \bigg(\frac43 - \frac{u}{2} \bigg)^2
    M(s, t) \,,
\label{mellin}
\eeq
where the Mellin variables are defined so that $s+t+u=0$. From the Mellin amplitude $M(s, t)$, we then define the Borel transform $A(S,T)$, via
\begin{equation}
 M(s, t) = \frac{1}{2R^6} \int_0^\infty d\beta e^{-\beta} \beta^5 A \left(\frac{s \beta}{2R^2},\frac{t \beta}{2R^2} \right).
\end{equation}
We take $A(S,T)$ as the working definition of the $AdS$ Virasoro-Shapiro amplitude. At leading order in the large $R$ expansion, this transform implements the flat space limit introduced in~\cite{Penedones:2010ue}, where $S,T$ have the interpretation of the usual Mandelstam variables in flat space. More generally $A(S,T)$ has an expansion in powers of $1/R^2$:
\begin{equation}
A(S,T) = A^{(0)}(S,T) + \frac{1}{R^2} A^{(1)}(S,T)  + \frac{1}{R^4} A^{(2)}(S,T) +\cdots \ ,
\end{equation}
with $A^{(0)}(S,T)$ being the Virasoro-Shapiro amplitude in flat space, and subsequent terms small curvature corrections around flat space.\footnote{The AdS radius is related to the 't Hooft coupling by $\sqrt{\lambda}=R^2/\alpha'$  and here we set $\alpha'=1$.} $A^{(0)}(S,T)$ can be obtained by a standard textbook computation and is naturally given in terms of a genus zero world-sheet integral. As shown in \cite{Alday:2023jdk,Alday:2023mvu}, higher-order curvature corrections, defined this way, also admit a representation in terms of genus zero world-sheet integrals. We will return to this point later. 

In this paper we consider $A(S,T)$ in the Regge (large $T$, finite $S$) limit. As familiar from Regge theory, this limit is dominated by the exchange of operators in the leading Regge trajectory. For the Mellin amplitude, we can then write
\begin{equation}
M(s,t) \simeq \sum_{J=0} C^2(J) M_{\tau,J}(s) t^J \ ,
\label{Reggedec}
\end{equation}
where the sum involves only operators in the leading Regge trajectory, and for each exchange we have only kept the leading contribution at large $t$. Here $\tau,J$ denote the twist and the spin of the exchanged operator, $C(J)$ denotes the OPE coefficient and $M_{\tau,J}(s)$ is presented in Appendix \ref{app:exchanges}. More precisely, one should first perform the sum over the spin, and then take the large $t$ limit. Since the leading contribution arises from the Regge spin $J=J^*$, it is a valid approximation to consider the leading large $t$ asymptotics of each exchange.\footnote{See \cite{Banerjee:2024ibt} for a concise review on Regge theory.}  

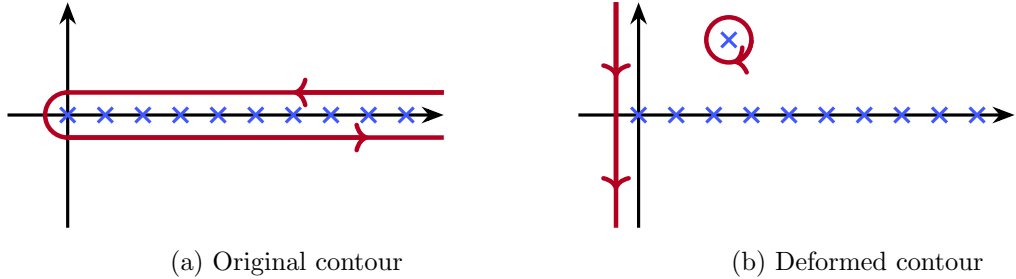
\begin{figure}
    \centering
    \begin{subfigure}[c]{0.48\textwidth}
        \begin{tikzpicture}
            \draw[very thick] (-0.8,1.5) -- (0,1.5);
            \draw[line width=0.6mm,color=integration_contours] (0,1.5) circle (0.3);
            \draw[color=white,fill=white] (0,0) -- (0,3) -- (2,3) -- (2,0) -- (0,0);
            \draw[-Stealth,very thick] (0,0) -- (0,3);
            \draw[-Stealth,very thick] (0,1.5) -- (5,1.5);
            \draw[line width=0.6mm,color=integration_contours] (-0.03,1.8) -- (5,1.8);
            \draw[line width=0.6mm,color=integration_contours] (-0.03,1.2) -- (5,1.2);
            \draw[->,color=integration_contours,line width=0.6mm] (5,1.8) -- (3,1.8);
            \draw[->,color=integration_contours,line width=0.6mm] (0,1.2) -- (4,1.2);
            
            \draw[color=graph_poles,very thick] (-0.1,1.4) -- (0.1,1.6);
            \draw[color=graph_poles,very thick] (-0.1,1.6) -- (0.1,1.4);

            \draw[color=graph_poles,very thick] (0.4,1.4) -- (0.6,1.6);
            \draw[color=graph_poles,very thick] (0.4,1.6) -- (0.6,1.4);

            \draw[color=graph_poles,very thick] (0.9,1.4) -- (1.1,1.6);
            \draw[color=graph_poles,very thick] (0.9,1.6) -- (1.1,1.4);

            \draw[color=graph_poles,very thick] (1.4,1.4) -- (1.6,1.6);
            \draw[color=graph_poles,very thick] (1.4,1.6) -- (1.6,1.4);

            \draw[color=graph_poles,very thick] (1.9,1.4) -- (2.1,1.6);
            \draw[color=graph_poles,very thick] (1.9,1.6) -- (2.1,1.4);

            \draw[color=graph_poles,very thick] (2.4,1.4) -- (2.6,1.6);
            \draw[color=graph_poles,very thick] (2.4,1.6) -- (2.6,1.4);

            \draw[color=graph_poles,very thick] (2.9,1.4) -- (3.1,1.6);
            \draw[color=graph_poles,very thick] (2.9,1.6) -- (3.1,1.4);

            \draw[color=graph_poles,very thick] (3.4,1.4) -- (3.6,1.6);
            \draw[color=graph_poles,very thick] (3.4,1.6) -- (3.6,1.4);

            \draw[color=graph_poles,very thick] (3.9,1.4) -- (4.1,1.6);
            \draw[color=graph_poles,very thick] (3.9,1.6) -- (4.1,1.4);

            \draw[color=graph_poles,very thick] (4.4,1.4) -- (4.6,1.6);
            \draw[color=graph_poles,very thick] (4.4,1.6) -- (4.6,1.4);
        \end{tikzpicture}\caption{Original contour}\label{fig:hankel_contour}
    \end{subfigure}
    \begin{subfigure}[c]{0.48\textwidth}
        \begin{tikzpicture}
            \draw[-Stealth,very thick] (0,0) -- (0,3);
            \draw[-Stealth,very thick] (-0.8,1.5) -- (5,1.5);
            
            \draw[line width=0.6mm,color=integration_contours] (-0.3,0) -- (-0.3,3);
            \draw[->,line width=0.6mm,color=integration_contours] (-0.3,3) -- (-0.3,0.5);
            \draw[->,line width=0.6mm,color=integration_contours] (-0.3,3) -- (-0.3,2);

            \draw[color=graph_poles,very thick] (1.1,2.4) -- (1.3,2.6);
            \draw[color=graph_poles,very thick] (1.3,2.4) -- (1.1,2.6);
            \draw[line width=0.6mm,color=integration_contours] (1.2,2.5) circle(0.3);
            \draw[->,line width=0.6mm,color=integration_contours] (1.5,2.5) arc [radius=0.3,start angle=0, end angle=-80];

            \draw[color=graph_poles,very thick] (-0.1,1.4) -- (0.1,1.6);
            \draw[color=graph_poles,very thick] (-0.1,1.6) -- (0.1,1.4);

            \draw[color=graph_poles,very thick] (0.4,1.4) -- (0.6,1.6);
            \draw[color=graph_poles,very thick] (0.4,1.6) -- (0.6,1.4);

            \draw[color=graph_poles,very thick] (0.9,1.4) -- (1.1,1.6);
            \draw[color=graph_poles,very thick] (0.9,1.6) -- (1.1,1.4);

            \draw[color=graph_poles,very thick] (1.4,1.4) -- (1.6,1.6);
            \draw[color=graph_poles,very thick] (1.4,1.6) -- (1.6,1.4);

            \draw[color=graph_poles,very thick] (1.9,1.4) -- (2.1,1.6);
            \draw[color=graph_poles,very thick] (1.9,1.6) -- (2.1,1.4);

            \draw[color=graph_poles,very thick] (2.4,1.4) -- (2.6,1.6);
            \draw[color=graph_poles,very thick] (2.4,1.6) -- (2.6,1.4);

            \draw[color=graph_poles,very thick] (2.9,1.4) -- (3.1,1.6);
            \draw[color=graph_poles,very thick] (2.9,1.6) -- (3.1,1.4);

            \draw[color=graph_poles,very thick] (3.4,1.4) -- (3.6,1.6);
            \draw[color=graph_poles,very thick] (3.4,1.6) -- (3.6,1.4);

            \draw[color=graph_poles,very thick] (3.9,1.4) -- (4.1,1.6);
            \draw[color=graph_poles,very thick] (3.9,1.6) -- (4.1,1.4);

            \draw[color=graph_poles,very thick] (4.4,1.4) -- (4.6,1.6);
            \draw[color=graph_poles,very thick] (4.4,1.6) -- (4.6,1.4);
            
        \end{tikzpicture}\caption{Deformed contour}\label{fig:deformed_contour}
    \end{subfigure}
    \caption{Integration contours in the complex-$J$ plane. The original contour reproduces the sum over spins. The deformed contour, closed to the right, picks up the original poles, as well as Regge poles, of which the residues are subtracted to keep the same result.}\label{integrationContours}
\end{figure}

The expansion (\ref{Reggedec}) has a corresponding expression after Borel transform:
\begin{equation}
A(S,T) \simeq \sum_{J=0} C^2(J) A_{\tau,J}(S) T^J \ ,
\end{equation}
where
\begin{equation}
M_{\tau,J}(s) = \frac{1}{2 R^6} \int_0^\infty d\beta e^{-\beta} \beta^5 A_{\tau,J} \left(\frac{s \beta}{2R^2}\right) \left( \frac{\beta}{2R^2}  \right)^J.
\end{equation}
We will be interested in expansions around the flat space limit. It is convenient to introduce the rescaled Mellin variable $\hat s = \frac{s}{2 R^2}$, which is kept finite as we take the flat space limit. Furthermore, we introduce
\begin{equation}
\left(\tau -\frac{4}{3}\right) \left(2 J+\tau +\frac{16}{3}\right) = m^2 R^2.
\end{equation}
The `mass' $m$ is also finite in the flat space limit.\footnote{Recall that at strong coupling $\tau \sim \lambda^{1/4} \sim R$.} $M_{\tau,J}(\hat s)$ satisfies a difference equation which we write in the form
\begin{equation}
{\cal D} M_{\tau,J}(\hat s) = M_c(\tau,J) \ ,
\end{equation}
with $M_c(\tau,J)$ independent of $\hat s$, and the difference operator can be written as
\begin{equation}
{\cal D} =(-4 U_+ + m^2) + \frac{1}{R^2} \sum_{n=0}^\infty \frac{4(-1)^n}{R^{2n}\Gamma(n+1)} \left( \frac{U_0^2+\left (\frac{8 (n+2)}{3}-1 \right)U_0}{(n+1)(n+2)}+\frac{16}{9} \right) U_-^n \ .
\label{Doperator}
\end{equation}
See Appendix~\ref{app:transforms} for the representation of the operators $U_{\pm},U_0$ acting on functions of $\hat s$. This relation in turns implies an analogous relation for $A_{\tau,J}(S)$:
\begin{equation}
{\cal D} A_{\tau,J}(S) = A_c(\tau,J) \ ,
\label{relA}
\end{equation}
where ${\cal D}$ takes exactly the same form as in (\ref{Doperator}) but now the operators $U_0,U_{\pm}$ act on functions of $S$. 
Next, we will apply the ideas of Regge theory to $A(S,T)$. We first write the sum over the spin $J$ as a contour integral:
\begin{equation}
A(S,T) = \int \frac{dJ}{2 i} \frac{1+(-1)^J}{2 \sin (\pi J)} C^2(J) A_{\tau,J}(S) T^J \ .
\end{equation}
In principle, we should sum over all trajectories, but as already mentioned, only the leading Regge trajectory, with $\tau=\tau(J)$, will contribute in the Regge limit. The contour integral picks the poles at $J=0,2,\cdots$. Note that we think of $\tau(J)$ and $C^2(J)$ as analytic functions of $J$. Then, we deform the contour, as in figure \ref{integrationContours}. The deformed contour, parallel to the imaginary axis, picks up residues from Regge poles, which must be subtracted from this new integral to give $A(S,T)$. In the large $T$ limit, the factor of $T^J$ causes the pole with the largest real value of $J$—which corresponds to the leading Regge trajectory—to provide the dominant contribution to the Borel amplitude.

The analytic structure of $A_{\tau,J}(S)$ is more complicated than in flat space. In particular, in a $1/R$ expansion, it contains poles of higher and higher order. The locations and residues of these poles are determined, in terms of $m(J)^2$, entirely by the  relation (\ref{relA}) that $A_{\tau,J}(S)$ satisfies. In a $1/R$ expansion, this relation becomes a recursion relation and can be solved order by order. To any given order in $1/R$, $A_{\tau,J}(S)$ contains poles of the form
\begin{equation}
A^{(k)}(S) = \frac{r_k(J)}{x^k},~~~x = -4S+m(J)^2 \ ,
\end{equation}
where $k$ is a positive integer number. In the $J-$plane, this leads to a $k$-th order pole at $J=J^*(S)$, where $J^*$ is implicitly given by the solution of
\begin{equation}
m(J^*)^2 = 4 S \ .
\end{equation}
Which contribution does this give in the Regge limit? In the $J$-plane, $-4S+m(J)^2$ has a simple zero at $J=J^*$, so that we can write
\begin{equation}
-4S+m(J)^2 \equiv (J-J^*)\beta(J) \ .
\end{equation}
Deforming the contour and using Cauchy's theorem, we see that the contribution from the pole $A^{(k)}(S)$ to the Regge limit is
\begin{equation}
\int \frac{dJ}{2 i} \frac{1+(-1)^J}{2 \sin (\pi J)} C^2(J)  \frac{r_k(m^2)}{(J-J^*)^k \beta(J)^k } T^J = \left. -\pi \frac{\partial_J^{k-1}}{\Gamma(k)} \frac{1+(-1)^J}{2 \sin (\pi J)} C^2(J)  \frac{r_k(J)}{\beta(J)^k } T^J \right|_{J=J^*} \ .
\end{equation}
In other words, for every pole of order $k$ in $A_{\tau,J}(S)$ we can write down its contribution as an operator 
\begin{equation}
A_{\tau,J}(S) \sim \frac{r_k(J)}{x^k} \to \frac{\partial_J^{k-1}}{\Gamma(k)} \frac{r_k(J)}{\beta(J)^k } 
\end{equation}
acting on functions of $J$. Note that the derivatives act on everything on the right. We can then formally define the operators
\begin{equation}
y=  \partial_J \frac{1}{\beta(J)},~~:y^{n}: \, =  \partial^n_J \frac{1}{\beta(J)^n} \ .
\end{equation}
Note the `normal ordered' $:y^2: \neq y y$. Rather, $:y^n:$ is defined so that all the derivative operators $\partial_J$ are placed on the left, and then they act on everything on their right. The result in the Regge limit is then given by
\begin{equation}
A_{Regge}(S,T) =-\pi :\mathcal R(y): \left. \frac{1+(-1)^J}{2 \sin (\pi J)} C^2(J)  \frac{1}{\beta(J)} T^J \right|_{J=J^*(S)} \ ,
\end{equation}
where $\mathcal R(y)$ is the following integral transform of $A(x)$:
\begin{equation}
A(x) = \int_0^\infty e^{-y x} \mathcal R(y) dy \ .
\end{equation}
This integral transform basically implements Cauchy's theorem. We can write down the final answer as follows. We define the normalised function $\hat{\mathcal{R}}(y)$ such that it satisfies the relation
\begin{equation}
{\cal D} \hat{\mathcal{R}}(y) = \delta(y) \ ,
\label{Rhatrel}
\end{equation}
where ${\cal D}$ has the same formal expression as before:
\begin{equation}
{\cal D} =(-4 U_+ + m^2)+\frac{1}{R^2}\sum_{n=0}^\infty \frac{4(-1)^n}{R^{2n}\Gamma(n+1)} \left( \frac{U_0^2+\left (\frac{8 (n+2)}{3}-1 \right)U_0}{(n+1)(n+2)}+\frac{16}{9} \right) U_-^n \ ,
\end{equation}
but now the operators $U_{\pm},U_0$ are defined as operators acting on functions of $y$ by
\begin{eqnarray}
U_0 = (m^2-\partial_y)y \ ,~~~U_+ = \frac{1}{4} (m^2-\partial_y) \ ,~~~U_- = -4 q y+4(m^2-\partial_y)y^2 \ .
\end{eqnarray}
In a large $R$ expansion:
\begin{equation}
\hat{\mathcal{R}}(y)  = \theta(y) + \cdots \ ,
\end{equation}
with $ \theta(y)$ the Heaviside step function. As an operator, $\theta(y)$ acts as the identity operator (since $y \geq 0$ is the range the integral transform is defined). Our final formula for the AdS Virasoro-Shapiro amplitude in the Regge limit is then
\begin{equation}
A_{Regge}(S,T) = :\hat{\mathcal{R}}(y): \left. \frac{1+(-1)^J}{2 \sin (\pi J)}{\cal C}(J)  \frac{1}{\beta(J)} T^J \right|_{J=J^*(S)} \ ,
\label{finalRegge}
\end{equation}
with ${\cal C}(J)$ a rescaled OPE coefficient,
\begin{equation}
{\cal C}(J) = - \frac{\pi 2^{J+1}R^{2J+4}}{\Gamma(J+6)} M_c(\tau,J) \, C^2(J) \ ,
\end{equation}
where $M_c(\tau,J)$ can be found in Appendix~\ref{app:exchanges} and $\tau=\tau(J)$ is the twist of the operators in the leading Regge trajectory.

\subsection{Small curvature expansion}

Let us compute explicitly all the ingredients of the expression (\ref{finalRegge}) in a $1/R$ expansion, for the first few orders. Let us start with the CFT data for the operators in the leading Regge trajectory. Their twist is given by \cite{Gromov:2011bz}
\begin{equation}
\tau(J) = \sqrt{2(J+2)} R -J-2+\frac{1}{R}\frac{3 J^2+10J+16}{4\sqrt{2(J+2)}} + \cdots \ .
\end{equation}
From this, we can determine $m^2$ and 
\begin{eqnarray}
 \beta(J)&=&2+\frac{3 J+6 S-16}{6 R^2}+\cdots \ , \\
 J^*(S) &=& 2 (S-1)+\frac{-9 S^2+33 S-10}{9
    R^2} + \cdots \ .
 \end{eqnarray}
 The OPE coefficients can be read off from \cite{Alday:2023mvu}. We have
\begin{equation}
 {\cal C}(J)=-\frac{4 \pi }{\Gamma
   \left(\frac{J+4}{2}\right)^2}+\frac{1}{R^2} \frac{2
   \pi  (-4 (J+2) \zeta (3)+7
   J+4)}{(J+2) \Gamma
   \left(\frac{J}{2}+1\right)^2} + \cdots \ .
\end{equation}
Finally, solving (\ref{Rhatrel}) in a $1/R$ expansion, and for $y>0$, we find
\begin{equation}
\hat{\mathcal{R}}(y)  = 1 - \frac{1}{R^2} \frac{2}{9}(3 m^4 y^3+6 m^2 y^2+2y)+ \cdots \ .
\end{equation}
With these ingredients we find

\begin{equation}
\begin{split}
A_{Regge}(S,T) &=A^{(0)}_{Regge}(S,T)+ \frac{1}{R^2}A^{(1)}_{Regge}(S,T)+ \cdots \\
&= A^{(0)}_{Regge}(S,T) \left(1+ \frac{1}{R^2}\hat A^{(1)}_{Regge}(S,T)+ \cdots \right) \ ,
\end{split}
\end{equation}
with 
\begin{equation}
A^{(0)}_{Regge}(S,T)= e^{i \pi S} \frac{\Gamma (-S)}{\Gamma (S+1)} T^{2 S-2} \ , 
\end{equation}
the flat space result in the Regge limit (where we have assumed $T$ has a small negative imaginary part) and for example
\begin{eqnarray*}
\hat A^{(1)}_{Regge}(S,T) &=& -\frac{4 S^2}{3} \log ^3 T + \frac{2}{3} S \left (-3 i \pi  S+3 \pi  S \cot (\pi  S)+6 S \psi ^{(0)}(S)-2 \right ) \log^2 T +\\
& & S^2 \left(2 i \pi ^2 \cot (\pi  S)+4 i \pi   \psi ^{(0)}(S) -1 -2 \pi ^2 \cot ^2(\pi  S)-4  \psi ^{(0)}(S)^2+2 \psi ^{(1)}(S)  \right. \\
& & \left. -4 \pi  \cot (\pi  S) \psi ^{(0)}(S)-\frac{4 i \pi }{3S}+\frac{11}{3 S}+\frac{4}{3 S} \pi  \cot (\pi  S)+\frac{8 \psi ^{(0)}(S)}{3 S}\right) \log T +\\
& &  \frac{1}{6} i \pi  S^2\left(\frac{11}{S}+\frac{4 \pi  \cot (\pi  S)}{S}-6 \pi ^2 \csc ^2(\pi  S)-12 \psi ^{(0)}(S)^2+\frac{8 \psi ^{(0)}(S)}{S}+6 \psi ^{(1)}(S) \right. \\
& & \left. -12 \pi  \cot (\pi  S) \psi ^{(0)}(S)+4 \pi ^2-3\right) +
2 S^2 \zeta (3)-\frac{7 S^2}{2}-\frac{2}{3} \pi ^3 S^2 \cot (\pi  S)+\frac{1}{2} \pi  S^2 \cot (\pi  S) \\
& & +\frac{4}{3} S^2 \psi ^{(0)}(S)^3-2 \pi ^2 S^2 \psi ^{(0)}(S)+S^2 \psi ^{(0)}(S)-2 S^2 \psi ^{(1)}(S) \psi ^{(0)}(S)+\frac{S^2 \psi ^{(2)}(S)}{3}\\
& & +\pi ^3 S^2 \cot (\pi  S) \csc ^2(\pi  S)+2 \pi  S^2 \cot (\pi  S) \psi ^{(0)}(S)^2-\pi  S^2 \cot (\pi  S) \psi ^{(1)}(S)\\
& &+2 \pi ^2 S^2 \csc ^2(\pi  S) \psi ^{(0)}(S)+\frac{2 \pi ^2 S}{3}+\frac{5 S}{2}+\frac{2}{3 S}-\frac{11}{6} \pi  S \cot (\pi  S)-\frac{2}{3} \pi ^2 S \csc ^2(\pi  S)\\
& &-\frac{4 S \psi ^{(0)}(S)^2}{3}-\frac{11 S \psi ^{(0)}(S)}{3}+\frac{2 S \psi ^{(1)}(S)}{3}-\frac{4}{3} \pi  S \cot (\pi  S) \psi ^{(0)}(S)-\frac{11}{6} \ ,
\end{eqnarray*}
which is quite complicated, but fully explicit. Note that the terms proportional to $\log^3 T, \log^2 T$ are relatively simple. 
$A^{(2)}_{Regge}(S)$ contains terms proportional to $ \log ^6 T, \log^5 T,\cdots$, and so on, which can also be explicitly computed (but are very involved). Higher order terms can be computed in terms of the CFT data of operators in the leading Regge trajectory. To each order in $1/R$, the result is complicated but completely explicit. Actually, to any order in $1/R$ we can write the result in terms of derivatives acting on the Regge limit of the flat space Virasoro-Shapiro amplitude. For instance,
\begin{equation}
A^{(1)}_{Regge}(S,T)= \left( P^{(2)}_{3}(S) \partial_S^3 +P^{(2)}_{2}(S) \partial_S^2+P^{(2)}_{1}(S) \partial_S+P^{(2)}_{0}(S) \right)A^{(0)}_{Regge}(S,T) \ ,
\end{equation}
with the second order polynomials given by
\begin{equation}
P^{(2)}_{3}(S)=-\frac{S^2}{6},~~P^{(2)}_{2}(S)=-\frac{4}{3}S,~~P^{(2)}_{1}(S)=\frac{11}{6}S-\frac{S^2}{2}-\frac{7}{3},~~P^{(2)}_0(S) = 2 S^2 \zeta (3)-\frac{7}{2}S^2+\frac{3}{2} S+\frac{11}{6}
\end{equation}
More generally, we obtain
\begin{equation}
\label{Reggestructure}
A^{(k)}_{Regge}(S,T)= \left( P^{(2k)}_{3k}(S) \partial_S^{3k} +\cdots+P^{(2k)}_{0}(S) \right)A^{(0)}_{Regge}(S,T) \ .
\end{equation}
From the expression (\ref{finalRegge}) we can also infer some results to all orders in $1/R$. More precisely, to order $1/R^{2k}$ we get terms proportional to $\log^{3k} T, \log^{3k-1} T,\cdots$. We can then resum all leading logs, all subleading logs, and so on. To compute this explicitly, we need to compute $\hat{\mathcal{R}}(y)$ in a limit of large $y,R^2$, with $y^3/R^2$ fixed. This can be easily done by keeping a finite number of terms in the sum in (\ref{Rhatrel}).  For the first orders we obtain
\begin{equation}
\hat{\mathcal{R}}(y) = \sum_{k=0} \frac{(-1)^k}{R^{2k}} \left( \frac{2 m^4}{3} \right)^k \frac{1}{\Gamma(k+1)} \left(y^{3k} + \frac{2 k (11-6k)}{5} \frac{y^{3k-1}}{m^2}+ \cdots   \right) \ .
\label{Rhatleading}
\end{equation}
With this at hand, we obtain, for the resummation of the leading logs,
\begin{equation}
A^{LL}_{Regge}(S,T) =A^{(0)}_{Regge}(S,T)  \times e^{-\frac{4}{3} \frac{S^2}{R^2} \log^3 T},
\end{equation}
so that the leading logs exponentiate. This makes contact with the high energy results of \cite{Alday:2023pzu}, and indeed, both results perfectly agree in the regime in which they can be compared. The limit considered in this paper, however, is much richer. For the subleading poles we obtain
\begin{equation}
\begin{split}
A^{SL}_{Regge}(S,T) = &A^{(0)}_{Regge}(S,T) \times \frac{1}{R^2} e^{-\frac{4}{3} \frac{S^2}{R^2} \log^3 T} \\
& \times \log^2 T \left(\frac{16 S^3 \log^3T}{5 R^2}+2 \pi  S^2 \cot (\pi S)+4 S^2 \psi ^{(0)}(S)-\frac{4 S}{3}-2 i \pi  S^2\right).
\end{split}
\end{equation}
To order $1/R^4$, this agrees with the explicit results obtained above. This expression, however, provides all order results in the limit of large $R,T$ with $\log^3T/R^2$ fixed. 

\section{World-sheet perspective}

The Regge limit of the AdS VS amplitude can also be computed from the world-sheet representation found in \cite{Alday:2023mvu}. In a large $R$ expansion, the amplitude takes the form
\begin{equation}
A(S,T) = \int d^2z |z|^{-2S-2}|1-z|^{-2T-2} G(S,T,z) \ ,
\end{equation}
with $G(S,T,z) = \sum_k 1/R^{2k} G^{(k)}(S,T,z)$ and 
\begin{equation}
G^{(0)}(S,T,z) = \frac{1}{3}\left(\frac{1}{U^2} + \frac{|z|^2}{S^2}+ \frac{|1-z|^2}{T^2} \right)
\end{equation}
reproducing the usual VS amplitude in flat space.\footnote{This choice is equivalent to the one in the introduction, but more symmetric-looking.} At an arbitrary order $1/R^{2k}$, $G^{(k)}(S,T,z)$ is built from single-valued multiple-polylogarithms of weight $3k$ and rational functions of $S,T$ of the form $P^{(2k)}(S,T)/U^2$ and crossing symmetric combinations, where $P^{(2k)}(S,T)$ are homogenous polynomials of degree $2k$. 

We are interested in the Regge, large $T$, finite $S$, limit. At the level of the world-sheet integrand, the relevant limit is that of small $z,\bar z$ and large $T$, with $T z, T \bar z $ fixed. We can then define
\begin{equation}
\hat G(S,T,z)= \lim_{\epsilon \to 0} \frac{1}{\epsilon^2} G \left (S,\frac{T}{\epsilon},\epsilon z \right) \ . 
\end{equation}
We assume the integrand remains finite as we take $\epsilon \to 0$. This is true for the results obtained in \cite{Alday:2023jdk,Alday:2023mvu}, provided the ambiguities that integrate to zero are chosen appropriately. Of course the final result does not depend on the choice of such ambiguities, but for this particular choice one can take the Regge limit at the level of the integrand. In the Regge limit, we then obtain
\begin{equation}
A_{Regge}(S,T) = \int d^2z |z|^{-2S-2}|1-z|^{-2T-2} \hat G(S,T,z) \ .
\end{equation}
The structure of $\hat G(S,T,z)$ in a $1/R$ expansion is relatively simple.  At each order in $1/R$, $\hat G(S,T,z)$ is given by a polynomial in $z,\bar z$ and $\log |z|$. More precisely 
\begin{equation}
\hat G^{(k)}(S,T,z) = \frac{S^{2k}}{T^2} h^{(k)}_0(\log |z|) +  \frac{S^{2k-1}}{T} h^{(k)}_1(z,\bar z, \log |z|) + \cdots + \frac{T^{2k}}{S^2} h^{(k)}_{2k+2}(z,\bar z, \log |z|) \ ,
\label{Ghatstructure}
\end{equation}
where $h^{(k)}_{q}(z,\bar z, \log |z|)$ is a polynomial of degree $3k$ in $ \log |z|$ and a homogenous symmetric polynomial of degree $q$ in $z,\bar z$. For example, we can explicitly compute $\hat G(S,T,z)$ from the results in \cite{Alday:2023mvu}. To leading order we find
\begin{equation}
\hat G^{(0)}(S,T,z)= \frac{2}{3} \frac{1}{T^2} + \frac{1}{3} \frac{|z|^2}{S^2} \ ,
\end{equation}
while for the first curvature correction
\begin{eqnarray}
\hat G^{(1)}(S,T,z)= & &\frac{2 S^2-2 S T (z+\bar z)+5 T^2 z \bar z}{18 T^2} \log^3|z|^2 + \frac{(z+\bar z) (T (z+\bar z)-2 S)}{6 T} \log^2|z|^2 \nonumber \\
&& + \frac{1}{4} \left(z^2+\bar z^2\right) \log |z|^2 - \frac{1}{4} \left(5 z^2+ 5 \bar z^2 +z \bar z (4-8 \zeta (3))\right) \ .
\end{eqnarray}
The second order integrand in the Regge limit $\hat G^{(2)}(S,T,z)$ can also be computed from the results in \cite{Alday:2023mvu}, but the final answer is very lengthy and not particularly illuminating. 

In general, in order to compute the integral from a given integrand, we can use the result 
\begin{equation}
I^{mn}(S,T)= \int d^2z |z|^{-2S-2}|1-z|^{-2T-2} z^m \bar z^n = \frac{\Gamma(m-S)\Gamma(-T)\Gamma(1-U-n)}{\Gamma(S+1-n)\Gamma(T+1)\Gamma(U+m)} 
\label{VZresult}
\end{equation}
of \cite{Vanhove:2018elu}, with the following Regge behaviour
\begin{equation}
I^{mn}_{Regge}(S,T) =e^{i \pi S} \frac{\Gamma (-S)}{\Gamma (S+1)} T^{2S}\times \frac{\Gamma (S+1)^2}{\Gamma (-m+S+1) \Gamma (-n+S+1)} T^{-m-n} \ ,
\end{equation}
where we have assumed $T$ is large with a small, negative imaginary part and we have pulled out the large $T$ result for $m=n=0$, relevant for Virasoro-Shapiro in flat space. Note furthermore that the Virasoro-Shapiro amplitude in flat space is given by $A^{(0)}(S,T) = U^{-2}  I^{00}(S,T)$. Insertions of $\log |z|^2$ in the integrand can be computed in terms of derivatives:
\begin{equation}
\int d^2z |z|^{-2S-2}|1-z|^{-2T-2} z^m \bar z^n \log^p|z|^2 = (-1)^p \ \partial_S^p \ I^{mn}(S,T) \ .
\end{equation}
With these results at hand, we can compute $A_{Regge}(S,T)$ to order $1/R^4$ from the world-sheet perspective. The results are in full agreement with what we found in the previous section. 
 
\subsection{World-sheet solution} 

In the following, we reverse our logic and look for integrands that lead to the correct result in the Regge limit, to all orders in $1/R$. For a given solution, let's say to order $1/R^{2k}$, note that we always have the freedom to add insertions $K^{(k)}(S,T,z)$ in the `kernel' such that they integrate to zero:
\begin{equation} 
\int d^2z |z|^{-2S-2}|1-z|^{-2T-2} K^{(k)}(S,T,z)=0 \ .
\end{equation} 
Given the structure (\ref{Reggestructure}), we look for integrands such that
\begin{equation}
\frac{1}{T^2} \int d^2z |z|^{-2S-2}|1-z|^{-2T-2} F^{(k)}(S,T,z) =\left( P^{(2k)}_{3k}(S) \partial_S^{3k} +\cdots+P^{(2k)}_{0}(S) \right)A^{(0)}_{Regge}(S) \ .
\end{equation}
It turns out that it is always possible to find a solution with the structure (\ref{Ghatstructure}), namely\footnote{Indeed, the insertion of positive powers of $z,\bar z$ produces polynomials in $S$ times the flat space Virasoro-Shapiro amplitude in the Regge limit, see (\ref{VZresult}), so that the choice below leads to an overcomplete basis of polynomials of degree $2k$ in $S$, upon integration, multiplying each power of $\log T$.}
\begin{equation}
F^{(k)}(S,T,z) =S^{2k} f^{(k)}_0(\log |z|) + S^{2k-1} T f^{(k)}_1(z,\bar z, \log |z|) + \cdots + \frac{T^{2k+2}}{S^2} f^{(k)}_{2k+2}(z,\bar z, \log |z|) \ ,
\end{equation}
where the functions $f^{(k)}_q(z,\bar z,\log |z|)$ are homogenous polynomials of degree $3k$ in $\log |z|$, the single-valued logarithm, as a consequence of the number of derivatives in (\ref{Reggestructure}). This structure is fully consistent, and provides independent evidence, for the structure proposed in \cite{Alday:2022xwz,Alday:2023jdk,Alday:2023mvu}. Note that these polynomials are not fully fixed, due to the possibility of adding kernels, as explained above. 

In order to find an integrand systematically, let's instead fix the kernel ambiguity by considering the following family of integrands:
\begin{equation}
\frac{1}{T^2} \int d^2 z |z|^{-J-4} |1-z|^{-2T-2} F\left(S,-\log|z| \right) \simeq -F(S,\partial_J) e^{i \pi \frac{J}{2}} \frac{\Gamma(-1-J/2)}{\Gamma(2+J/2)} T^J \ ,
\end{equation}
where $F\left(S,-\log|z| \right)$ is a generic insertion that depends only on $\log|z|$ ({\it i.e.} it has no powers of $z,\bar z$ in a small $z,\bar z$ expansion), and the above equality is true in the Regge limit. In this expression, $J$ is simply a parameter and $S$ a spectator. Note that the factor $|z|^{-J-4}$ coincides with the usual factor $|z|^{-2S-2}$ at leading order in $1/R$ upon setting $J=J^*(S)$, which we will do momentarily. With this family of insertions, we get the following functional equation in order to reproduce the correct result in the Regge limit, namely equation (\ref{finalRegge}):
\begin{equation}
\left.-F(S,\partial_J) e^{i \pi \frac{J}{2}} \frac{\Gamma(-1-J/2)}{\Gamma(2+J/2)} T^J \right|_{J=J^*(S)}= :\hat{\mathcal{R}}\left(\partial_J \frac{1}{\beta(J)}\right): \left. \frac{1+(-1)^J}{2 \sin (\pi J)}{\cal C}(J)  \frac{1}{\beta(J)} T^J \right|_{J=J^*(S)} \ .
\label{functionaleq}
\end{equation}
On both sides of this equation, one should first take derivatives w.r.t. $J$ and then set $J=J^*(S)$. This should be seen as an equation for the insertion $F(S,\partial_J)$ in terms of $\hat{\mathcal{R}}(y)$, which can be explicitly computed to any order, and the CFT data for the operators in the leading Regge trajectory. To any order in $1/R$, this equation admits a unique solution. Indeed, to order $1/R^{2k}$, $\hat{\mathcal{R}}(y)$ is a polynomial of degree $3k$. Hence, $F(S,\partial_J)$ is a polynomial of degree $3k$ in the `variable' $\partial_J$:
 \begin{equation}
 \left. F(S,\partial_J) \right|_{1/R^{2k}} = f_{3k}(S) \partial_J^{3k} +\cdots+ f_1(S) \partial_J +f_0(S) \ .
\end{equation}
The coefficients of this polynomial are then uniquely fixed by matching the powers of $\log T$ on both sides of the functional equation (\ref{functionaleq}). For clarity of notation, let us introduce the following
\begin{equation}
G(\partial_J)\equiv :\hat{\mathcal{R}}\left(\partial_J \frac{1}{\beta(J)}\right): \ ,~~~ -e^{i \pi \frac{J}{2}} \frac{\Gamma(-1-J/2)}{\Gamma(2+J/2)} \equiv h(J) \ ,~~~\frac{1+(-1)^J}{2 \sin (\pi J)}{\cal C}(J)  \frac{1}{\beta(J)} \equiv g(J) \ .
\end{equation}
Hence we are looking for $F(S,\partial_J)$ such that
\begin{equation}
 \left.F(S,\partial_J) h(J) T^J \right|_{J=J^*(S)}=  \left.G(\partial_J) g(J)T^J\right|_{J=J^*(S)} \ .
\end{equation}
But because this equation is valid for all $T$, we can write an operator solution:
\begin{equation}
F(S,\partial_J) = \left. G(\partial_J) \frac{g(J)}{h(J)} \right|_{J=J^*(S)} \ .
\end{equation}
On the r.h.s., all the derivatives $\partial_J$ are on the left (due to the normal ordering in $:\hat{\mathcal{R}}(\partial_J \frac{1}{\beta(J)}):$). At any order in $1/R$, there is a finite number of such derivatives. We then move all derivatives to the right (as operators), and then set $J=J^*(S)$. This gives $F(S,\partial_J)$.

As an example, we can compute $F(S,\partial_J)$ to order $1/R^2$. We can do this by plugging all the ingredients, given in the previous section, to this order. We find
\begin{equation}
F(S,\partial_J) = 1+ \frac{1}{R^2} \left(-\frac{4}{3} S^2 \partial_J^3 -\frac{16}{3} S \partial_J^2 -\frac{32}{9}  \partial_J +2  \zeta (3)S^2-\frac{7}{2}S^2+\frac{3 }{2}S+\frac{11}{6} \right)+ \cdots \ .
\end{equation}
The resulting insertion does not have the structure  (\ref{Ghatstructure}). The reason for this is that we fixed the kernel ambiguity by forbidding positive powers of $z,\bar z$ in a small $z,\bar z$ expansion. Relaxing this condition we can make the insertion consistent with (\ref{Ghatstructure}), as seen above. Both solutions differ by a kernel that integrates to zero of course. 

As another example, let's find the insertion $F(S,-\log |z|)$ that reproduces the leading $\log^{3k} T$ to each order $1/R^{2k}$. In this approximation, we can set $h(J)$ and $g(J)$ to their flat space values:
\begin{equation}
 \frac{g(J)}{h(J)} = 1 + O\left(\frac{1}{R^2}\right) \ ,
\end{equation}
together with $\beta(J) = 2+ \cdots$, $m^2(J^*)=4S+\cdots$. From (\ref{Rhatleading}), keeping only the leading order terms at each order in $1/R$ we then find
\begin{equation}
F\left( \partial_J \right) = \ :\hat{\mathcal{R}} \left (\frac{1}{2}\partial_J \right): \ \simeq e^{-\frac{4 S^2}{3 R^2} \partial_J^3} \ .
\end{equation}
The corresponding world-sheet integral is then
\begin{equation}
\frac{1}{T^2} \int d^2 z |z|^{-J^*(S)-4} |1-z|^{-2T-2} e^{\frac{4 S^2}{3 R^2} \log^3|z|} \ , 
\end{equation}
where in this approximation we could take $J^*(S)=2S-2$ of course. To this solution, we can add an integrand that integrates to zero. More precisely, combining this result with the proposal in~\cite{Alday:2023mvu} leads to\footnote{In particular, note that for the leading logs at each order in $1/R$ the series in $z,\bar z$ truncates to second order and does not contain $z^2+\bar z^2$. This is a consequence of the specific structure of the proposal in~\cite{Alday:2023mvu}.}
\begin{eqnarray*}
A^{LL}_{Regge}(S,T) =&& \frac{1}{T^2} \int d^2 z |z|^{-2S-2} |1-z|^{-2T-2} e^{\frac{4 S^2}{3 R^2} \log^3|z|}\times \\
&&\times \left(f_0\left(\frac{S^2}{R^2}\log^3|z| \right) + (z+\bar z)\frac{T}{S} f_1\left(\frac{S^2}{R^2}\log^3|z| \right)+z \bar z \frac{T^2}{S^2}f_2\left(\frac{S^2}{R^2}\log^3|z| \right) \right) \ ,
\end{eqnarray*}
where requiring the correct leading order logs to all orders in $1/R$, in the Regge limit, implies
\begin{equation}
f_0\left(\frac{S^2}{R^2}\log^3|z| \right)+2f_1\left(\frac{S^2}{R^2}\log^3|z| \right)+f_2\left(\frac{S^2}{R^2}\log^3|z| \right)=1 \ .
\end{equation}
This is a prediction for the integrand, to all orders in $1/R$, in the limit of large $R$, small $z$ with $\log^3|z|/R^2$ fixed.  

\section{Conclusions}

In this paper, we have computed the AdS Virasoro-Shapiro amplitude in the Regge (large $T$, finite $S$) limit, in terms of the CFT data of intermediate operators in the leading Regge trajectory. Our result (\ref{finalRegge}) depends on three ingredients: the dimension of the operators in the leading Regge trajectory, through $m^2(J)$, which determines $J^*(S)$ and the Regge slope $\beta(J)$, the OPE coefficient ${\cal C}(J)$ with which these operators enter, and the operator $\hat{\mathcal{R}}(y)$, which can be seen as encoding the effect of the curvature of the background. To any order in $1/R$ the answer can be written down explicitly, and takes the form of derivatives acting on the flat space Virasoro-Shapiro amplitude in the Regge limit. 

We then worked out the implications at the level of the world-sheet representation, finding families of solutions, for the world-sheet integral, that lead to the correct Regge behaviour to all orders in the curvature expansion. The structure of the integrand is fully consistent with, and provides strong evidence for, the proposal put forward in \cite{Alday:2023mvu}. In particular, the Regge limit makes manifest the single-valuedness of the integrand around $z=0$. 

The solutions we have found for the integrand, have large families of ambiguities given by integrands that integrate to zero. Similar ambiguities were also present in \cite{Alday:2023jdk,Alday:2023mvu}. While at present it is not clear how to fully fix or understand these ambiguities, we note that for a specific choice (that fixes them partially), one can take the Regge limit at the level of the integrand first. 

An open question is how to impose the correct Regge limit together with single-valuedness at $z=1$. This should settle the question of how constraining the CFT data of the leading twist operators is in constructing the AdS Virasoro-Shapiro amplitude. The results of this paper are the first steps in this direction.  

The results of this paper connect very directly the AdS Virasoro-Shapiro amplitude to the impressive developments in integrability \cite{Gromov:2011de,Basso:2011rs,Gromov:2011bz}, via the CFT data of leading twist operators in planar ${\cal N}=4$ SYM. It would be fascinating to develop a {\it quantum spectral curve} for the AdS Virasoro-Shapiro amplitude.

It would be very interesting to reproduce our results for the AdS Virasoro-Shapiro amplitude by a direct string theory computation. Recently there has been progress in developing string field theory in the presence of RR-backgrounds, see \cite{Cho:2023mhw,Chostrings}. It should be possible to use this framework to compute  the AdS 
Virasoro-Shapiro amplitude in a small curvature expansion.

\section*{Acknowledgements} 

We would like to thank T. Hansen for very useful discussions. The work of LFA is partially supported by the STFC grant ST/T000864/1. The work of CV is supported by the Fonds de recherche du Qu\'ebec, secteur Nature et technologies, through its Doctoral Training Scholarships program, scholarship number 349150. The work of X.Z. is supported by the NSFC Grant No. 12275273, funds from Chinese Academy of Sciences, University of Chinese Academy of Sciences, the Kavli Institute for Theoretical Sciences, and also by the Xiaomi Foundation. LFA, CV and XZ acknowledge support from the the European Research Council (ERC) under the European Union's Horizon 2020 research and innovation programme (grant agreement No 787185).

\appendix

\section{Details on exchanges}\label{app:exchanges}

In our conventions, the Mellin amplitude corresponding to the exchange of an operator of spin $J$ and twist $\tau$ is given by\footnote{Because this corresponds to an exchange diagram for the {\it reduced} Mellin amplitude, one should set the dimension of the external operators to $\Delta_{external}=4$ when comparing this to an usual exchange Mellin amplitude.}
\begin{equation}
M_{\tau,J}(s,t)= \sum_{m=0}^\infty \frac{\mathcal{Q}_{J, m}^{\tau+4, d=4}(t)}{s+\frac{4}{3}-\tau-2m} +P_{J-1}(s,t) \ ,
\end{equation}
where
\begin{align}
\mathcal{Q}_{J, m}^{\tau, d}(t) \equiv -\frac{2^{3 J+2 \tau -2} \Gamma \left(J+\frac{\tau }{2}-\frac{1}{2}\right) \Gamma \left(J+\frac{\tau }{2}+\frac{1}{2}\right) \Gamma \left(-\frac{d}{2}+J+\tau +1\right) Q_{J, m}^{\tau, d}(t)}{\pi  \Gamma (m+1) \Gamma \left(J+\frac{\tau }{2}\right)^2 \Gamma (J+\tau -1) \Gamma \left(-m-\frac{\tau }{2}+4\right)^2 \Gamma \left(-\frac{d}{2}+J+m+\tau +1\right)} \ ,
\end{align}
with $Q_{J, m}^{\tau, d}(t)$ a Mack polynomial \cite{Mack:2009mi}. Mack polynomials are in general very complicated, but in this paper, we are only interested in the leading large $t$ behaviour for which
\begin{equation}
Q_{J, m}^{\tau, d}(t) = t^J + \cdots \ .
\end{equation}
Furthermore, in this limit, we can also ignore regular terms present in the exchange, given by a polynomial $P_{J-1}(s,t)$ of degree $J-1$, since this grows at most as $t^{J-1}$ in the large $t$ limit. In this limit, we can compute explicitly the exchange amplitude $M_{\tau,J}(s,t) =M_{\tau,J}(s) t^J+\cdots $, but we are more interested in the difference equation it satisfies:
\begin{equation}
\left(s-\frac{8}{3}\right)^2 M_{\tau,J}(s-2) - \left(s^2 + 2J s+ \frac{20}{3}s -R^2 m^2 \right) M_{\tau,J}(s) = M_c \ ,
\end{equation}
where we have defined
\begin{equation}
\left(\tau -\frac{4}{3}\right) \left(2 J+\tau +\frac{16}{3}\right) = m^2 R^2 \ .
\end{equation}
This relation follows from the equation of motion identity satisfied by the exchange Witten diagram. When expanding around the flat space limit, $m^2$ will be kept fixed as we take $R$ large. $M_c$ is a complicated but explicit quantity, independent of $s$:
\begin{equation}
M_c(\tau,J) = \frac{ 2^{3 J+2 \tau +7} \Gamma (J+6) \Gamma \left(J+\frac{\tau }{2}+\frac{3}{2}\right) \Gamma \left(J+\frac{\tau }{2}+\frac{5}{2}\right)}{\pi \Gamma \left(2-\frac{\tau }{2}\right)^2 \Gamma \left(J+\frac{\tau }{2}+2\right)^2 \Gamma \left(J+\frac{\tau }{2}+4\right)^2} \ .
\end{equation}

\section{Integral transforms}\label{app:transforms}
In the body of the paper, we introduced several integral transforms. It will be useful to understand how different operators act on different spaces. First, consider the space of functions $f(\hat s)$ and $F(S)$. They are related by
\begin{equation}
f(\hat s) = \kappa \int_{0}^\infty d\beta e^{-\beta} \beta^{-q-1} F(\hat s \beta) \ ,
\end{equation}
where $\kappa$ is a constant, and to make a connection with the transform used in the body of the paper $\hat s= \frac{s}{2R^2}$, where $s$ is the usual Mellin variable and $q=-6-J$. We introduce three operators $U_0$ and $U_{\pm}$ such that their action is local in the three spaces we are interested in. In functions of $\hat s$ they act as
\begin{equation}
U_0 = \hat s \partial_{\hat s} \ ,~~~ U_+= -q \hat s + \hat s^2 \partial_{\hat s} \ ,~~~ U_- = \partial_{\hat s} \ ,
\end{equation}
from where we can check the following commutation relations
\begin{equation}
[U_0,U_{\pm}] = \pm U_{\pm} \ ,~~~[U_+,U_-] = -2 U_0+q \ .
\end{equation}
Given the integral transform above, we can then work their action on functions of $S$, to obtain
\begin{equation}
U_0 = S \partial_{S} \ ,~~~ U_+= S \ ,~~~ U_- = -q \partial_S + S \partial^2_S \ .
\end{equation}
The third integral transform is simply the Laplace transform in $(x,y)$, where $x=-4S+m^2$, given by
\begin{equation}
A(x) = \int_0^\infty dy e^{-y x} \mathcal R(y) \ .
\end{equation}
The representation of $U_0,U_{\pm}$ as operators acting on functions of $y$ is then
\begin{eqnarray}
U_0 = (m^2-\partial_y)y \ ,~~~U_+ = \frac{1}{4} (m^2-\partial_y) \ ,~~~U_- = -4 q y+4(m^2-\partial_y)y^2 \ .
\end{eqnarray}

\section{Scalar propagator on $AdS$ around flat space}\label{app:propagator}
The Virasoro-Shapiro amplitude on AdS $A(S,T)$ has an interesting analytic structure in a $1/R$ expansion. In particular, to order $1/R^{2k}$, it has poles of order $3k+1$ at the masses of the exchanged particles. In this appendix, we show that this structure is mimicked by the AdS scalar propagator around flat space when written in `momentum space'. 

Let us analyse first the Mellin amplitude for a scalar exchange in a CFT in $d-$dimensions, which satisfies the following relation:
\begin{equation}
(\hat s R^2-\Delta_0)^2 {\cal M}_{\text{ex}}\left (\hat s - \frac{1}{R^2} \right) - \left( \hat s^2 R^4- \frac{d}{2} \hat s R^2 - \frac{m^2}{4} \right){\cal M}_{\text{ex}}(\hat s)  = {\cal M}_c \ ,
\end{equation}
where $\tau$ is the dimension/twist of the exchanged operator, $\Delta_0$ is the dimension of the external operators and we have introduced $\tau(\tau-d)=m^2 R^2$. Here $\hat s=\frac{s}{2R^2}$ is the rescaled Mellin variable. In the flat space limit, we take $R$ large with fixed $m^2,\Delta,\hat s$. The Borel transform takes exactly the same form as in Appendix~\ref{app:transforms}, but now $2q=d-2\Delta_0$:
\begin{equation}
{\cal M}_{\text{ex}}(\hat s) = \kappa \int_{0}^\infty d\beta e^{-\beta} \beta^{-q-1} A_{\text{ex}}(\hat s \beta) \ .
\end{equation}
The Borel transform then satisfies the relation
\begin{equation}
{\cal D} A_{\text{ex}}(S) = A_{\text{contact}} \ ,
\label{Arelap}
\end{equation}
with $A_{\text{contact}}$ independent of $S$ and 
\begin{equation}
{\cal D} =(-4 U_+ + m^2) + \frac{1}{R^2} \sum_{n=0}^\infty \frac{4(-1)^n}{\Gamma(n+1)R^{2n}}\left( \frac{U_0^2+\left((n+2)\Delta_0-1 \right)U_0}{(n+1)(n+2)} + \Delta_0^2 \right) U_-^n \ .
\end{equation}
The operators $U_{\pm},U_0$ acting on functions of $S$ were defined in the previous appendix and we recall them here:
\begin{equation}
U_0 = S \partial_{S} \ ,~~~ U_+= S \ ,~~~ U_- = -q \partial_S + S \partial^2_S \ .
\end{equation}
The analytic structure of $A_{\text{ex}}(S)$ follows from (\ref{Arelap}), which can be easily solved in a $1/R$ expansion.  At order $1/R^{2k}$ we get poles of order $3k+1$ and lower. In order to compute them, it suffices to consider the first term in the sum in ${\cal D} $. After an appropriate rescaling, we obtain
\begin{equation}
\left( (-4 S + m^2) + 2 S^2 \partial_S^2 + \cdots \right) A_{\text{ex}}(S)=1 \ ,
\end{equation}
which reproduces the towers of poles of order $3k+1$ to order $1/R^{2k}$.

Let us now turn to the propagator. The AdS bulk-to-bulk scalar propagator for a particle of dimension $\Delta$ is given by
\begin{equation} 
G(X,Y) = C_\Delta \zeta^{-\Delta} ~_2F_1\left( \Delta, \Delta - \frac{d}{2}+\frac{1}{2}; 2\Delta-d+1;- \frac{4}{\zeta} \right) \ ,
\end{equation}
where the chordal distance is
\begin{equation}
\zeta= \frac{(\vec X-\vec Y)^2-(X_0-Y_0)^2}{R^2} \ .
\end{equation}
Here $X_0=\sqrt{R^2+\vec X^2},Y_0=\sqrt{R^2+\vec Y^2}$. Next, we introduce
\begin{equation}
\zeta = \frac{r^2}{R^2} \ ,~~~\Delta(\Delta-d)=m^2 R^2 \ ,
\end{equation}
so that the expansion around flat space is defined as an expansion around large $R$, keeping $r,m$ finite. The simplest way to compute the expansion of the propagator around flat space is by considering the equation it satisfies:
\begin{equation} 
\left( \left( \partial_r^2 + \frac{d}{r} \partial_r -m^2 \right) +\frac{1}{R^2} \frac{r}{4} \left((2d+1)\partial_r + r \partial^2_r \right) \right) G(r) =\delta(r) \ .
\label{propagatoreq}
\end{equation}
The operator acting on $G(r)$ can be written in terms of the Laplacian in $d+1$ (flat) dimensions and the dilatation operator
\begin{equation}
\nabla^2= \partial_r^2 + \frac{d}{r} \partial_r,~~~D= r \partial_r  \ ,
\end{equation}
where we are interpreting $r$ as the radial coordinate in the flat $d+1$ dimensions. In terms of these
\begin{equation} 
\left( \nabla^2 -m^2 +\frac{1}{R^2}\left( \frac{d}{2(d-1)}r^2 \nabla^2 -\frac{d+1}{4(d-1)}D^2 \right) \right) G(r) =\delta(r) \ .
\label{propagatoreq}
\end{equation}
Expanding $G(r) = G^{(0)}(r) + \frac{1}{R^2} G^{(1)}(r)+ \cdots$, we see that at leading order we get the usual equation for the propagator in flat space, in position space. Next, we go to `momentum' space, which we simply define as follows
\begin{equation}
G(p_1,p_2) = \int d^{d+1} x d^{d+1} y e^{i p_1 \cdot x}e^{i p_2 \cdot y} \tilde G(x-y) \ ,
\end{equation}
where we interpret $r=|x-y|$ as the distance between the points $x,y$, which parametrise the flat $d+1$ dimensional space. Writing $z_i=x_i-y_i$ and $w_i=x_i+y_i$, then $ \nabla^2 = \partial_{z_i} \partial_{z_i}$ and $D=z_i \partial_{z_i}$. One can define the momentum space propagator in AdS in different ways~\cite{Gadde:2022ghy}, but this will not affect our discussion on the leading order poles. Integrating over $w$ gives rise to the delta function $\delta(p_1+p_2)$. Then, introducing the `Mandelstam variable' $\tilde s= -p^2$, with $p_1=-p_2=p$, the equation for the propagator in momentum space becomes
\begin{equation} 
\left( \tilde s -\mu^2 +\frac{1}{R^2}\left( \tilde s^2 \partial^2_{\tilde s} +2 \tilde s  \partial_{\tilde s}  + \frac{1-d^2}{4} \right) \right) \tilde G(\tilde s) =1 \ .
\end{equation}
At leading order in a $1/R$ expansion, we get the familiar expression 
\begin{equation}
\tilde G(\tilde s) = \frac{1}{\tilde s -\mu^2}  + \cdots \ ,
\end{equation}
while at order $1/R^{2k}$ we get poles of order $3k+1$. Note that the leading order poles are determined by the second order derivative in the subleading term
\begin{equation} 
\left( \tilde s -m^2 +\frac{1}{R^2}\left( \tilde s^2 \partial^2_{\tilde s} +\cdots \right) \right) \tilde G(\tilde s) =1 \ ,
\end{equation}
with a structure analogous to the poles in the exchange Mellin amplitude in Borel space, and hence the AdS Virasoro-Shapiro amplitude.

\bibliographystyle{jhep}
\bibliography{Bibliography}

\end{document}